\documentclass[prb,aps,twocolumn,showpacs,superscriptaddress]{revtex4}

\usepackage{graphicx}

\begin{document}

\title{Predicting the spin-lattice order of frustrated systems from first-principles}

\author{H. J. Xiang} 
\email{hxiang@fudan.edu.cn}
\affiliation{Key Laboratory of Computational Physical Sciences (Ministry of Education), and Department of Physics, Fudan
  University, Shanghai 200433, P. R. China}

\author{E. J. Kan}
\affiliation{Department of Applied Physics, Nanjing University of
  Science and Technology, Nanjing, Jiangsu 210094, P. R. China}

\author{Su-Huai Wei}
\affiliation{National Renewable Energy Laboratory, Golden, Colorado 80401, USA}

\author{M.-H. Whangbo}
\affiliation{Department of Chemistry, North Carolina State
  University, Raleigh, North Carolina 27695-8204, USA}

\author{X. G. Gong}
\affiliation{Key Laboratory of Computational Physical Sciences
  (Ministry of Education), and Department of Physics, Fudan
  University, Shanghai 200433, P. R. China}

\date{\today}

\begin{abstract}
  A novel general  method of describing the spin-lattice
  interactions 
  in magnetic solids was proposed in terms of first principles calculations. 
  The spin exchange and Dzyaloshinskii-Moriya interactions as well
  as their derivatives
  with respect to atomic displacements can be evaluated efficiently on
  the basis of
  density functional calculations for four ordered spin states.
  By taking into consideration the spin-spin interactions, the
  phonons, and  the coupling between them,  we show that
  the ground state structure of a representative spin-frustrated spinel, MgCr$_2$O$_4$, is tetragonally
  distorted, in agreement with experiments. However,
  our calculations find the lowest energy for the collinear
  spin ground state, in contrast to
  previously suggested non-collinear models.
\end{abstract}


\maketitle

\section{Introduction}
Spin frustrated systems \cite{Gardner2010,Balents2010,Saunders2008} have recently attracted considerable
attention because of their novel magnetic
properties. 
The geometrically frustrated spin lattice generally  leads to 
numerous degenerate spin configurations.
In principle, a strongly frustrated system mostly has no
long-range spin order, but the degeneracy can often be lifted by
the spin-lattice coupling; a symmetry-lowering lattice 
distortion gives rise to a long-range magnetic order at low
temperature.
In the emerging field of multiferroics, \cite{Cheong2007} it was found that 
many frustrated magnets (such as RMnO$_3$ \cite{Kimura2003} and
RMn$_2$O$_5$ \cite{Hur2004}) display a large magnetoelectric coupling.
Magnetic frustration combined
with a striking spin-lattice coupling appears to  cause
their multiferroic properties.
In particular, the exchange striction (a kind of spin-lattice
coupling  arising from the dependence of the symmetric exchange interactions
on the atomic positions)
is considered to induce the ferroelectricity in
some collinear antiferromagnets such as RMn$_2$O$_5$.
\cite{Chapon2004}  An alternative mechanism of multiferroicity
comes from the inverse effect of Dzyaloshinskii-Moriya (DM)
interaction \cite{Sergienko2006} (another kind of  spin-lattice
coupling arising from 
the dependence of antisymmetric exchange interactions
on  the atomic positions), which 
explains the ferroelectricity in many noncollinear spiral
magnets such as RMnO$_3$
\cite{Sergienko2006,Malashevich2008,Xiang2008} and
Ni$_3$V$_2$O$_8$. \cite{Lawes2005} A significant contribution
of the symmetric $\mathbf{S_i}\cdot \mathbf{S_j}$-type magnetostriction
to the ferroelectricity in RMnO$_3$ was also revealed by Mochizuki
{\it et al.}, who estimated the dependence of the nearest-neighbor (NN)
ferromagnetic (FM)
coupling in RMnO$_3$ on the Mn-O-Mn bond angle in an empirical way.
\cite{Mochizuki2010}
It is clear that a quantitative discription  of the spin-lattice
coupling is desirable for further study of frustrated magnets.

In their pioneering work, Fennie and Rabe \cite{Fennie2006}
developed 
a first principles method to
calculate the second order derivatives of the symmetric exchange
parameter and the spin-phonon
coupling parameter, and they
studied the influence of magnetic order on the optical
phonons of the geometrically frustrated spinel ZnCr$_2$O$_4$. 
However,  their work does not fully describe how the spins couple
to the lattice, \cite{Fennie2006} and how
to predict the
spin-lattice order in frustrated systems from first principles is unclear.
Here
in this work, we propose a general first principles 
description of spin-lattice coupling, which allows one to efficiently
evaluate the
symmetric and antisymmetric exchange interaction parameters and their 
first-order derivatives with respect to atom displacements. 
As a test of our method, we examined a representative frustrated 
system, the spinel MgCr$_2$O$_4$, to find that its ground-state
structure
is tetragonally distorted  with collinear antiferromagnetic (AFM) spin
configuration.

In section II, we will describe a method for computing the
spin-lattice coupling parameters. The approach for predicting the
spin-lattice ground state using the 
spin-lattice coupling parameters will be discussed in section
III. Then we will apply the methods to find the spin-lattice ground
state of MgCr$_2$O$_4$ in section IV. Finally, we will summarize our
work in section V.

\section{Quantitative description of the spin-lattice coupling}
\subsection{Method for computing the symmetric exchange parameters}
Consider a classical  Heisenberg spin system,
whose energy
can be written as $E=E_0  +  E_{spin}$, where
$E_{spin}=\sum_{<i,j>} J_{ij} {\mathbf S_i} \cdot {\mathbf S_j} $ is
the spin exchange term with $|{\mathbf S_i}|=S$, and $E_0$ is the energy due
to other interactions (e.g., the lattice elastic energy), which
depends on the atom displacements  but not the spin orientations.
The spin exchange interactions 
are short range interactions and become negligible when the distance
between the spin sites $i$ and $j$ is 
longer than a certain critical 
distance $R_c$.
Given a supercell large enough so that any spin
site has no interaction with its neighboring cells, we extract
the exchange interactions as follows. 
Without loss of generality, consider a particular exchange
interaction $J_{12}$ between spin sites 1 and 2. The spin
Hamiltonian can be written as: 
\begin{equation}
  E_{spin}= J_{12} {\mathbf S_1} \cdot {\mathbf S_2} + {\mathbf S_1}
  \cdot {\mathbf K_1} + {\mathbf S_2} \cdot {\mathbf  K_2} + E_{other},
\end{equation}
where ${\mathbf K_1}=\sum_{i \ne 1,2} J_{1i}  {\mathbf S_i}$,
${\mathbf  K_2}=\sum_{i \ne
  1,2} J_{2i}  {\mathbf S_i}$, $E_{other}=\sum_{i,j \ne 1,2} J_{ij}
{\mathbf S_i} \cdot {\mathbf S_j}$.
It should be noted that ${\mathbf K_1}$, ${\mathbf K_2}$, and
$E_{other}$ do not depend on the spin directions of sites 1 and 2.
Consider the following four collinear spin states (with $z$ as the spin
quantization axis):
(1) $S_1^z = S$, $S_2^z = S$;
(2) $S_1^z = S$, $S_2^z = -S$;
(3) $S_1^z = -S$, $S_2^z = S$;
(4) $S_1^z = -S$, $S_2^z = -S$.
In these four spin states, the spin orientations for the spin sites
other than 1 and 2 are the same. One can easily show that the four
states have the following energy expressions:
\begin{equation}
  \begin{array}{c}
    E_1 = E_0 + E_{other} + J_{12} S^2+ K_1 S+ K_2S, \\
    E_2 = E_0 + E_{other} - J_{12} S^2+ K_1S  -K_2S, \\
    E_3 = E_0 + E_{other} - J_{12}S^2 - K_1S + K_2S, \\
    E_4 = E_0 + E_{other} + J_{12}S^2 - K_1S - K_2S.
  \end{array}
\end{equation}
Thus, $J_{12}$ is extracted by the formula:
\begin{equation}
  J_{12} = \frac{E_1+E_4-E_2-E_3}{4S^2}.
\end{equation}
The total energies of the four states can be calculated using
density functional theory (DFT).
The method described above is a
kind of mapping analysis \cite {Whangbo2003} that has been used widely to extract the
exchange parameters. 
In the usual mapping process, one usually considers the spin states
with small supercells to reduce computational demand. 
Here the use of a large supercell has the advantage that the
extraction of a particular exchange interaction is independent of
other exchange interactions and all total energies are computed
using the same supercell. In principles, the accuracy of the
exchange parameters from this method is only limited by the
predefined cutoff distance $R_c$, which can be checked by increasing
the supercell size.

\subsection{Method for computing the derivatives of the symmetric exchange parameters}
The above approach considered how to evaluate the exchange
parameters for a given structure.
We now examine
the dependence of the total energy on the atom displacements:    $E({\mathbf u_1},...,{\mathbf u_n},{\mathbf
  S_1},...,{\mathbf S_m})=E_0({\mathbf      U}) + \sum_{<i,j>}
J_{ij}({\mathbf U})
{\mathbf S_i} \cdot {\mathbf S_j}$, 
where ${\mathbf U} = ({\mathbf u_1},...,{\mathbf u_n})$ and ${\mathbf u_k}$ ($1 \le k \le n$) denote the 
displacements of atom $k$ from a reference structure,
$n$ and $m$ are the total number of atoms and total number of spin
sites in the supercell, respectively.
By taking the partial derivative of the above equation with respect to $u_{k\alpha}$, we obtain: 
\begin{equation}
  \frac {\partial E}{ \partial {u_{k\alpha}}} = \frac {\partial E_0}
        { \partial {u_{k\alpha}}} + \sum_{<i,j>} \frac {\partial J_{ij} }
        { \partial {u_{k\alpha}}} {\mathbf S_i} \cdot {\mathbf S_j}.
\end{equation}
In terms of the same four spin states used for extracting $J_{12}$,
the derivative of $J_{12}$ with respect to $u_{k\alpha}$ is found
as:
\begin{equation}
  \frac {\partial J_{12} } { \partial {u_{k\alpha}}} = \frac{1}{4S^2}(\frac
        {\partial E_1}  { \partial {u_{k\alpha}}} + \frac
        {\partial E_4} { \partial {u_{k\alpha}}}  - \frac {\partial
          E_2} { \partial {u_{k\alpha}}} - \frac {\partial E_3} {
          \partial {u_{k\alpha}}} ).
        \label{eq1}
\end{equation}
Here 
$-\frac {\partial E_i}  { \partial {u_{k\alpha}}}$ ($i=1,...,4$) is the force
acting on the atom $k$ along the $\alpha$ direction. The force can be computed
using the Hellmann-Feynman theorem  and is
readily available in many standard DFT schemes.
From Eq.~\ref{eq1}, we can see that the dependence of the exchange parameter
$J_{12}$ on all the atom displacements 
can be computed by performing four static total energy
calculations. 
This means that the calculation of first order derivative of
the exchange parameter does not require extra calculations if
one calculates the exchange parameter using our above method.  
Therefore, our new approach utilizing the Hellmann-Feynman forces  has a great computational advantage over the finite
difference method in which each $\frac {\partial J_{12} } { \partial
  {u_{k\alpha}}}$ requires several total energy calculations. 

\subsection{Methods for computing DM interaction parameters,
  single-ion anisotropy parameters, and their derivatives}
Our method for calculating the symmetric exchange parameter and its
derivative can be also extended to compute the antisymmetric DM interaction
parameter ( ${\mathbf D}$) and single-ion anisotropy (SIA) parameter ($A$) and their
derivatives. 
Let us describe the method of calculating the DM interaction parameter (vector $\mathbf{D_{12}}$) between spin site
$1$ and spin site $2$. Here, we calculate the three components 
$D_{12}^{x}$, $D_{12}^{y}$ and $D_{12}^{z}$ of
the DM vector separately for a general system, although
sometimes the direction of the DM vector can be determined by the crystal
symmetry. Without loss of  generality, let us focus on 
the calculation of $D_{12}^{z}$. 
We  consider the following four spin configurations in which 
the spins 1 and 2 are oriented along the $x-$ and $y-$axes, respectively: 
(1) ${\mathbf S_1} = (S,0,0)$, ${\mathbf S_2} = (0,S,0)$,
(2) ${\mathbf S_1} = (S,0,0)$, ${\mathbf S_2} = (0,-S,0)$,
(3) ${\mathbf S_1} = (-S,0,0)$, ${\mathbf S_2} = (0,S,0)$,
(4) ${\mathbf S_1} = (-S,0,0)$, ${\mathbf S_2} = (0,-S,0)$.
In these four spin configurations, the spins of all the other
spin sites are the same and are along the $z$ direction.  
The spin interaction energy for the four spin configurations can be written as:
\begin{equation}
  E_{spin}= D_{12}^{z} S_1^x S_2^y - S_1^x \sum_{i \ne 1,2} D_{1i}^y
  S_{i}^z + S_2^y \sum_{i \ne 1,2} D_{2i}^x
  S_{i}^z + E_{other}.
\end{equation}
As in the case of the symmetric exchange, 
we have
\begin{equation}
  \begin{array}{ccl}
     D_{12}^{z} &=& \frac{1}{4S^2} (E_1+E_4-E_2-E_3), \\
     \frac {\partial D_{12}^z } { \partial {u_{k\alpha}}} &=&
     \frac{1}{4S^2}(\frac
          {\partial E_1}  { \partial {u_{k\alpha}}} + \frac
          {\partial E_4} { \partial {u_{k\alpha}}}  - \frac {\partial
            E_2} { \partial {u_{k\alpha}}} - \frac {\partial E_3} {
            \partial {u_{k\alpha}}} ).
   \end{array}
 \end{equation}  
 Since the DM interaction is a consequence of spin-orbit coupling
 (SOC), 
 it is necessary that the energies of the four ordered spin states be
 determined 
 by DFT calculations with SOC effects taken into consideration. 
 The other two components $D_{12}^{x}$ and $D_{12}^{y}$  can be 
 computed in a similar manner. 
 This is 
 a general method to compute both the direction and
 the magnitude of a DM interaction parameter.
 For the calculation of the single-ion anisotropy parameter,  
 we consider the spin site 1.
 If the spin has an easy-axis (with local $z'$ axis) or easy-plane anisotropy, the
 single-ion anisotropy term  can be expressed as:
 $H_{sia}=A_1 S_{z'}^2$. 
 To evaluate $ A_1$, we consider the 
 four spin states in which the spin directions for site 1 are along
 $z'$,
 $-z'$, $x'$, and $-x'$ with the spins at all the other
 spin sites along the $y$ direction.
 One can easily find that $A_1 = \frac{E_1+E_2-E_3-E_4}{2S^2}$ and
 $\frac{\partial{A_1}}{ \partial {u_{k\alpha}}} =  \frac{1}{2S^2}
 (\frac           {\partial E_1}  { \partial {u_{k\alpha}}} + 
 \frac {\partial E_2} { \partial {u_{k\alpha}}}  - \frac {\partial
   E_3} { \partial {u_{k\alpha}}} - \frac {\partial E_4} { \partial
   {u_{k\alpha}}} )$. 
 Here the total energy calculations should include
 SOC effects because the single-ion anisotropy is a consequence of
 SOC.

\section{Prediction of the spin-lattice order}
It was shown that the spin-lattice coupling 
may lead to a distortion of the lattice to lower the
exchange energy
and relieve the frustration of a frustrated spin system (the
``spin-Teller'' effect \cite{Tchernyshyov2002}).
How the lattice distorts is determined by the balance
between the spin exchange and lattice elastic energies, and
depends on the parameters of the full spin-lattice coupled
Hamiltonian.
With the exchange parameters and their first order
derivatives with respect to the atom displacements in hand, one 
can predict its spin-lattice ground
state with a lower crystal symmetry.
At high temperature, a magnetically frustrated system is
usually in a disordered paramagnetic (PM) state with a high
symmetry. 
We now write the total energy of a frustrated system with the
PM state as a reference state:
\begin{equation}        
  E({\mathbf u_1},...,{\mathbf u_n},{\mathbf
    S_1},...,{\mathbf S_m})=E_{PM} + E_{ph} + E_{spin},
  \label{eq_tot}
\end{equation}
where $E_{ph}=1/2 \sum_{ij\alpha\beta} C_{ij}^{\alpha\beta}
u_{i\alpha}u_{j\beta}$ is the phonon Hamiltonian
($C_{ij}^{\alpha\beta}$ is the force constant), and $E_{spin}=
\sum_{<i,j>} [J_{ij}({\mathbf U}) {\mathbf S_i}
  \cdot {\mathbf S_j} + {\mathbf D_{ij}}({\mathbf U})\cdot ({\mathbf
    S_i} \times {\mathbf S_j})] +     \sum_{i} A_i({\mathbf U})
S_{i z'}^2$.       
Here $J_{ij}({\mathbf U})=J_{ij}^0+ \sum_{k\alpha}  \frac {\partial J_{ij} } { \partial {u_{k\alpha}}}
u_{k\alpha} $, $J_{ij}^0$ and $ \frac {\partial J_{ij} } { \partial
  {u_{k\alpha}}}$ are computed using the PM structure. We have
similar expressions for ${\mathbf D_{ij}}({\mathbf U})$ and
$A_i({\mathbf U})$.    The particular lattice distortion leading to the lowest energy for a
given spin configuration can be obtained  by solving the
following linear equations:
\begin{equation}
  \frac {\partial E } { \partial {u_{k\alpha}}} =  \sum_{j\beta} C_{kj}^{\alpha\beta}  u_{j\beta} +
  \frac {\partial E_{spin}} { \partial {u_{k\alpha}}} = 0
\label{eq_leq}
\end{equation}
where $ 1\le k \le n$ and $\alpha=1,2,3$. The lowest energy for a
given spin configuration can then be calculated by using
Eq.~\ref{eq_tot}. The spin-lattice ground state can be found by
comparing the energies of  different  spin configurations.

\section{Application to MgCr$_2$O$_4$}

\subsection{Computational details of DFT calculations}
Our total energy
calculations are based on the DFT plus the on-site
repulsion (U) method \cite{Liechtenstein1995} within the
local density approximation (LDA$+$U) on the basis of the projector
augmented
wave method \cite{PAW} encoded in the Vienna ab initio simulation
package \cite{VASP}.
We used  the on-site repulsion
$U = 3$ eV and the exchange parameter $J=0.9$ eV on Cr, which
reproduce the dominant features of the
photoemission and band gap data in sulfur Cr$^{3+}$ spinels
\cite{Fennie2006}.
The plane-wave cutoff energy was set to 400 eV.

\begin{figure}
  \includegraphics[width=8.0cm]{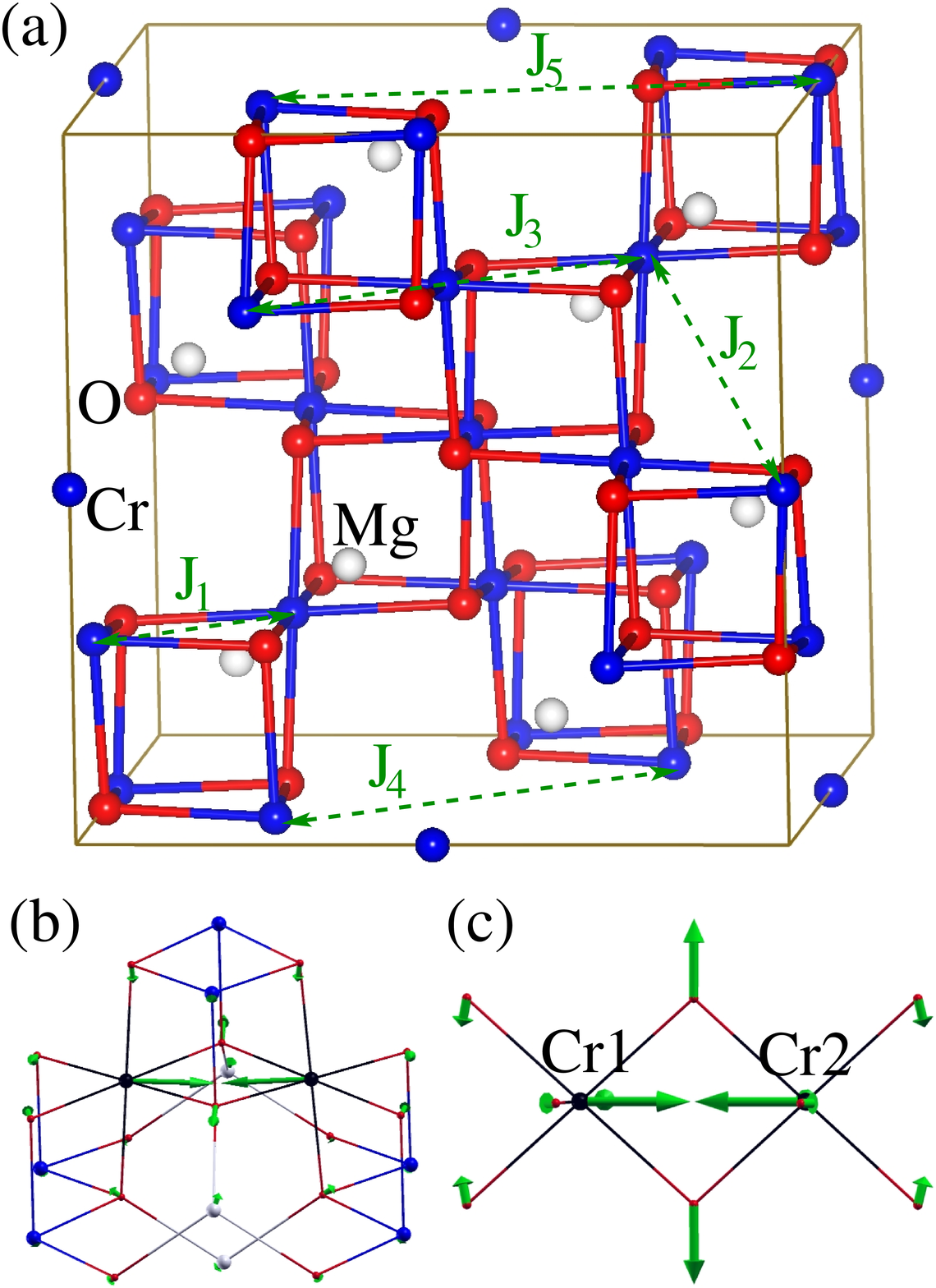}
  \caption{(Color online) (a) Structure of MgCr$_2$O$_4$ and
    exchange paths. (b) and (c) show the side and top views of the
    derivative of the NN exchange $J_1$ with respect to the atom displacements.}
  \label{fig1}
\end{figure}

\begin{figure}
  \includegraphics[width=7cm]{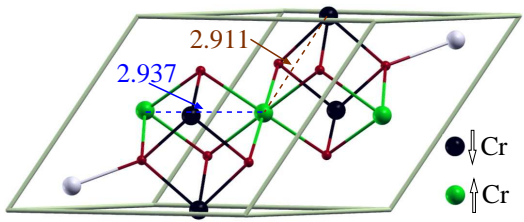}
  \caption{(Color online) The ground state of MgCr$_2$O$_4$ by
    considering the full spin-lattice coupled Hamiltonian. The
    numbers (in \AA) denote the Cr-Cr distances. The exchange
    parameter for the Cr-Cr pair with the 2.911
    \AA\ distance is 5.47 meV, while the exchange
    parameter for the Cr-Cr pair with the 2.937 
    \AA\ distance is 3.84 meV. }
  \label{fig2}
\end{figure}

\subsection{Exchange parameters and their derivatives}
We now apply the above method to calculate the exchange parameters and
their derivatives
of MgCr$_2$O$_4$ \cite{Ortega-San-Martin2008} to find its spin-lattice
coupled ground state.
MgCr$_2$O$_4$,  which crystallizes in a normal spinel
structure with the Cr$^{3+}$ ions ($S=3/2$) forming a pyrochlore
lattice [see Fig.~\ref{fig1}(a)], is strongly frustrated 
due to the strong 
AFM interactions between the NN  Cr$^{3+}$ spins. 
We consider first the MgCr$_2$O$_4$
structure optimized with the FM spin state. The optimized
lattice constant is 8.277\ \AA. We calculate all
symmetric spin exchange interactions up to the fourth  NN 
pairs [see Fig.~\ref{fig1}(a)] using a $2\times2 \times2$ supercell of
the MgCr$_2$O$_4$ conventional cubic cell;
the NN exchange $J_1$ within each Cr$_4$ tetrahedron and the farther NN
exchanges $J_2$, $J_3$, $J_4$, and $J_5$. 
We find that $J_1= 4.56$
meV, $J_2= 0.01$ meV, $J_3 = 0.26 $ meV, $J_4 = 0.08 $ meV, and
$J_5 = -0.01 $ meV.
The NN exchange $J_1$ is strongly AFM
while all next NN exchange interactions are almost negligible.

Using Eq.~\ref{eq1}, we calculate the derivatives of the 
exchange parameters
with respect to atom displacements using the optimized FM structure. Our results show the NN
exchange $J_1$ depends strongly on the positions  of those atoms
shown  in  Figs.~\ref{fig1}(b) and (c).
In particular, the largest derivative $|\frac {\partial J_{1} } {
  \partial {\mathbf{u_{k}}}}|$ of the exchange interaction
between two NN Cr ions  occurs when site $k$ is one of two Cr ions
(We hereafter refer to this vector as $\mathbf{J}_{1\mathrm{Cr}}^{'}$). 
When the two Cr ions move close to each other, the NN
exchange $J_1$ increases. The magnitude of the derivative is as
large as $43.40$
meV/\AA, which is close to the value (40.25 meV/\AA) extracted by the finite difference method.
This can be understood because when the
distance between two Cr ions becomes short, the direct overlap between the
$t_{2g}$ orbitals of the two Cr$^{3+}$ ($d^3$) ions becomes
stronger. 
The NN exchange $J_1$ also depends substantially on
the positions of the two bridging oxygen atoms
with its derivative approximately along the direction 
from the midpoint of the two NN Cr ions to the oxygen
ion (hereafter this vector is referred to as $\mathbf{J}_{1\mathrm{O}}^{'}$). 
We find the
direction of $\mathbf{J}_{1\mathrm{O}}^{'}$ is due to the fact that
the anti-bonding repulsion between O $p$ orbital and Cr $t_{2g}$
orbitals becomes weaker and 
the $t_{2g}$ orbitals of the two Cr$^{3+}$ ($d^3$) ions can have a
better overlap if the bridge O atoms move away from the Cr-Cr pair. 
Another possible explanation is that the increased Cr-O distance might
result in a smaller FM contribution, enhancing the overall AFM
coupling. 
The derivatives of other symmetric exchange parameters are found to
be vanishingly small.

It was suggested that the DM interaction might be relevant to the
spin-lattice order in a similar system. \cite{Ji2009}
As expected from the
symmetry analysis, the DM
vector for a Cr-Cr edge of each Cr$_4$ tetrahedron is
perpendicular to the
Cr-Cr bond and is parallel to the opposite edge of the
Cr$_4$  tetrahedron. Using our  method, we find the
magnitude of the DM
parameter to be 0.03 meV (0.7\% of $J_1$). 
We have checked that our method is accurate enough to predict
reliable DM vectors.
And the derivative of the DM
parameter with respect to the Cr ion position of the Cr-Cr
pair has the largest magnitude of 0.41 meV/\AA.  
Our calculations show that the Cr$^{3+}$ ion has an
easy plane anisotropy with the plane perpendicular to the
three-fold rotational axis $z'$.  The calculated SIA
parameter is $-0.05$ meV (1.1\% of $J_1$) and the
largest derivative of the SIA
is 0.18 meV/\AA. Our first principles calculation
established that, for MgCr$_2$O$_4$, the DM parameter, SIA
parameter, and theirs derivatives are negligible compared to NN
$J_1$.

\subsection{Spin-lattice ground state of MgCr$_2$O$_4$}
At high temperature, MgCr$_2$O$_4$ is paramagnetic (PM) with a cubic
symmetry. 
To simulate the PM  state, we use
a special quasirandom structure (SQS) \cite{Zunger1990} spin
configuration.
The magnetic unit cell of the SQS spin
configuration is four times as large as the chemical unit cell
of MgCr$_2$O$_4$. Then we fix the lattice constant to that (8.259 \AA) of the SQS
structure and relax the internal atomic coordinates with  the FM
spin configuration. In this way,
we can get approximate exchange parameters
$J_{ij}^0$ (4.80 meV),
their derivative $\frac {\partial J_{ij} } {\partial
  {u_{k\alpha}}} $ ($|\mathbf{J}_{1\mathrm{Cr}}^{'}|=49.11$
meV/\AA\ and $|\mathbf{J}_{1\mathrm{O}}^{'}|=25.15$
meV/\AA), and the force constants
$C_{ij}^{\alpha\beta}$ appropriate for the PM state. 
The differences between the force constants of the FM state and those
of the PM state are small because they are of second order
effect, and thus are neglected in this work.
All these
parameters are necessary in finding the spin-lattice coupled ground state
of MgCr$_2$O$_4$ by solving Eq.~\ref{eq_leq}.
The force constants are
obtained by finite difference as in phonon calculations by the
direct method. The force constants are considered fixed and
independent of the atomic displacement.

Our calculations show that the dominant exchange interaction in
MgCr$_2$O$_4$ is the NN AFM symmetric exchange interaction $J_1$. 
For the Heisenberg Hamiltonian with the NN  exchange interaction
$J_1$ on the pyrochlore lattice ($H_{NN}=\sum_{<\mathrm{NN}\, i,j>}
J_{1} {\mathbf                           S_i} \cdot {\mathbf S_j}$),
the degeneracy of the spin ground state is macroscopic:
If for any of the Cr tetrahedra, the sum of the four spins is
zero, then it is a spin ground state. \cite{Tchernyshyov2002}
We will use our calculated parameters (not only the exchange
parameter, but also its derivatives) and solve the full
spin-lattice coupled
Hamiltonian to determine the ground state of MgCr$_2$O$_4$.  
As a first step, we consider the case where the spin-lattice ground
state has the same size as 
the primitive chemical unit cell. In this case, we can easily generate
a spin configuration \cite{Tchernyshyov2002} that is one of the highly
degenerate spin ground states of the
Heisenberg  Hamiltonian $H_{NN}$. We generate two thousand of
such spin configurations and calculate the energy of the spin-lattice
coupled system, to find that the spin configuration shown in
Fig.~\ref{fig2} has the lowest energy and thus is the ground
state of the spin-lattice
coupled Hamiltonian. The spin state is collinear with two up spins
and two down spins. 
To confirm the above prediction from the model Hamiltonian analysis, 
we carry out DFT calculations to optimize both the lattice parameters
and the internal
coordinates of MgCr$_2$O$_4$ with the collinear AFM spin state. 
The relaxed (i.e., distorted) structure is calculated to have
a lower energy by 6.33 meV/Cr than the
unrelaxed structure with the same collinear AFM spin state. 
In the relaxed structure, the distance between NN spin up Cr$^{3+}$
ion and spin down Cr$^{3+}$ ion is smaller by 0.026 \AA\ than
that between NN Cr$^{3+}$ ions with the same spin direction.
The exchange parameters for the two different Cr-Cr exchange
interactions are 5.47 meV and 3.84 meV, respectively, to be compared
with the value (4.80 meV) for the undistorted structure.
This difference is due to the spin-lattice coupling, which makes
the exchange parameter between AFM (FM) coupled spins larger
(smaller). 
The relaxed MgCr$_2$O$_4$ structure is tetragonal (space group
$I4_1/amd$, No. 141) with $a= 5.873$
\AA\ and $c=8.160$ \AA. Experimentally,  the spinel
MgCr$_2$O$_4$ was found to undergo a sharp first order transition at
$T_N=12.4$ K from  a cubic paramagnetic phase (space group $Fd\bar{3}m$) to a tetragonal
antiferromagnetically ordered structure ($I4_1/amd$, $a =
5.8961$\AA\ and $c = 8.3211$ \AA\ at 10 K). \cite{Ortega-San-Martin2008} Our first principles
result thus confirms the $I4_1/amd$ space group of the ground
state of MgCr$_2$O$_4$ and $c<\sqrt{2}a$ below $T_N$. 
It should be noted that our work predicts a
collinear AFM  ground state with
the propagation vector $\mathbf{q}=(0,0,0)$ with respect to the
tetragonal lattice, in contrast to the previously
proposed non-collinear magnetic models. \cite{Ortega-San-Martin2008,Shaked1970}
This calls for further ultra-low temperature experiments to verify our prediction.

\section{Conclusion}
In summary, we proposed a general and efficient method to   
quantitatively describe spin-lattice coupling. 
This method allows one to evaluate each spin exchange (and DM
interaction) as well as its derivatives with respect to atom
displacements on the basis of DFT calculations for four ordered spin states.
By applying this method to the spin-frustrated spinel
MgCr$_2$O$_4$, we showed that it undergoes a structural transition from the cubic
to a tetragonal
structure with collinear AFM spin configuration.  
Our method provides an efficient first principles way of
describing the interplay between spin 
order and lattice distortion in frustrated magnetic systems.

Work at Fudan was partially supported by NSFC No. 11104038,
Pujiang plan, and Program for Professor of Special Appointment
(Eastern Scholar). Work at NREL was supported by U.S. DOE under
Contract No. DE-AC36-08GO28308, and that at NCSU U.S. DOE, under
Grant No. DE-FG02-86ER45259.




\begin{thebibliography}{99}
  
\bibitem{Gardner2010}J. S. Gardner, M. J. P. Gingras, and J, E. Greedan,
  Rev. Mod. Phys. {\bf 82}, 53 (2010).

\bibitem{Balents2010}L. Balents, Nature {\bf 464}, 199 (2010).

\bibitem{Saunders2008}T.E. Saunders and J.T. Chalker,Phys. Rev. B {\bf
  77}, 214438 (2008); 
  G.-W. Chern, C. J. Fennie, and O. Tchernyshyov, Phys. Rev. B {\bf 74}, 060405 (2006).
  
\bibitem{Cheong2007}S.-W. Cheong and M. Mostovoy, Nat. Mater. {\bf 6},
  13 (2007); R. Ramesh and N. Spaldin, Nat. Mater. {\bf 6}, 21 (2007).

\bibitem{Kimura2003}T. Kimura,
T. Goto, H. Shintani, K. Ishizaka, T. Arima, and Y. Tokura, Nature  {\bf 426}, 55 (2003)

\bibitem{Hur2004}N. Hur, S. Park, P. A. Sharma, J. S. Ahn, S. Guha,
  and S-W. Cheong, Nature {\bf 429}, 392 (2004).

\bibitem{Chapon2004}L. C. Chapon, G. R. Blake, M. J. Gutmann,
  S. Park, N. Hur, P. G. Radaelli, and S-W. Cheong,
  Phys. Rev. Lett. {\bf 93}, 177402 (2004).

\bibitem{Sergienko2006}I. A. Sergienko and E. Dagotto, Phys. Rev. B
  {\bf 73}, 094434 (2006).

\bibitem{Malashevich2008}A. Malashevich and D. Vanderbilt, Phys. Rev. Lett. {\bf 101}, 037210 (2008).

\bibitem{Xiang2008}H. J. Xiang, Su-Huai Wei, M.-H. Whangbo, and
  Juarez L. F. Da Silva,  
  Phys. Rev. Lett. {\bf 101}, 037209 (2008).  

\bibitem{Lawes2005}G. Lawes, A. B. Harris, T. Kimura, N. Rogado,
  R. J. Cava, A. Aharony, O. Entin-Wohlman, T. Yildrim, M. Kenzelmann,
  C. Broholm, and A. P. Ramirez, Phys. Rev. Lett. {\bf 95}, 087205 (2005).  

\bibitem{Mochizuki2010}M. Mochizuki, N. Furukawa, and N. Nagaosa,
  Phys. Rev. Lett. {\bf 105}, 037205 (2010).

\bibitem{Fennie2006}C. J. Fennie and K. M. Rabe, Phys. Rev. Lett. {\bf 96},
  205505 (2006).


\bibitem{Whangbo2003}M.-H. Whangbo, H.-J. Koo, and D. Dai
J. Solid  State Chem. {\bf 176}, 417 (2003). 

\bibitem{Tchernyshyov2002}O. Tchernyshyov, R. Moessner, and
  S. L. Sondhi,  Phys. Rev. Lett. {\bf 88}, 067203 (2002);
  Phys. Rev. B {\bf 66}, 064403 (2002).

\bibitem{Liechtenstein1995}A. I. Liechtenstein, V. I. Anisimov, and
  J. Zaanen, Phys. Rev. B  {\bf 52},  R5467 (1995).

\bibitem{PAW}P. E. Bl\"ochl, Phys. Rev. B {\bf 50}, 17953 (1994);
  G. Kresse and D. Joubert, {\it ibid}  {\bf 59}, 1758 (1999).

\bibitem{VASP}G. Kresse and J. Furthm\"uller, Comput. Mater. Sci. {\bf                                       
  6}, 15 (1996); Phys. Rev. B {\bf 54}, 11169 (1996).


\bibitem{Ortega-San-Martin2008}L. Ortega-San-Mart\'in, 
  A. J. Williams, C. D. Gordon, S, Klemme, and J. P. Attfield
  J. Phys.:
  Condens. Matter {\bf 20}, 104238 (2008).


\bibitem{Ji2009} S. Ji, S.-H. Lee, C. Broholm,
  T. Y. Koo, W. Ratcliff, S.-W. Cheong, and P. Zschack
  Phys. Rev. Lett. {\bf 103}, 037201 (2009).

\bibitem{Zunger1990}A. Zunger, S.-H. Wei, L. G. Ferreira, and James
  E. Bernard,  
  Phys. Rev. Lett. {\bf 65}, 353 (1990);
  S.-H. Wei, L. G. Ferreira, James E. Bernard, and A. Zunger,  
  Phys. Rev. B {\bf 42}, 9622 (1990).

\bibitem{Shaked1970}Shaked, J. M. Hastings, and L. M. Corliss,   Phys. Rev. B {\bf 1}, 3116 (1970).


\end{thebibliography}
\end{document}